\documentclass[12pt]{article}
\usepackage{ssi}
\usepackage{times}
\usepackage[dvips]{graphicx}
\usepackage{epsf}  % for embedding postscript figures
\usepackage{epsfig}  % for those that use epsfig instead
\usepackage{relsize}
\def\babar{\mbox{\slshape B\kern-0.1em{\smaller A}\kern-0.1em
    B\kern-0.1em{\smaller A\kern-0.2em R}}}
\newcommand{\psfiletwo}[3]{ % (arguments: full filenames, width/textwidth)
  \begin{minipage}{\linewidth}
    \parbox[b]{.49\linewidth}{%
      \begin{center}
        \setlength{\epsfxsize}{#3\linewidth}\leavevmode\epsfbox{#1}
      \end{center}
    }
    \hfill
    \parbox[b]{.49\linewidth}{%
      \begin{center}
        \setlength{\epsfxsize}{#3\linewidth}\leavevmode\epsfbox{#2}
      \end{center}
    }
  \end{minipage}
}

\begin{document}

\title{Measurements of Rare B Decays at \babar\ }

\author{Paul C. Bloom\thanks{Supported by DOE Contract DE-FG03-95ER40894.
%\thanks{Supported by DOE Contract xxx. 
% Footnotes in ten-point font.
\vskip 0.5in 
\noindent
\copyright\ 2002 by Paul C. Bloom.}\\ 
University of Colorado, Boulder \\
bloom@slac.stanford.edu\\[0.4cm]
%and \\[0.4cm]
%Jane Eyre\thanks{Supported by NSF Contract xxx.}\\ 
%Washington University, Seattle, Washington xxxxx \\[0.4cm]
Representing the \babar\ Collaboration
}

\maketitle
\begin{abstract}%
\baselineskip 16pt 

We present the results of searches for rare B meson decays.  The measurements 
use all or part of a data sample of about 88 million $\Upsilon(4S)\rightarrow 
B{\bar B}$ decays collected between 1999 and 2002 with the \babar\ 
detector at the PEP-II 
asymmetric energy B Factory at the Stanford Linear Accelerator Center.  
We study a variety of decays dominated 
by electromagnetic, electroweak and gluonic penguin transitions, and report 
measurements of branching fractions and other quantities of interest.

\end{abstract}

\section{Introduction}

Measurements\cite{babar_kstargamma,babar_rhogamma,babar_semibsg,babar_fullbsg,babar_kll,babar_ll,babar_knunu,babar_2gamma,babar_hphm,babar_hpi0,babar_hpi02,babar_2pi0,babar_phi,babar_omega,babar_etap,babar_3h} 
of rare B meson branching 
fractions have been performed using the 
\babar\cite{babarnim} detector.  B decays in which CKM favored amplitudes are
suppressed or forbidden 
are sensitive to penguin amplitudes and hence to possible 
non-Standard Model effects arising from new particles participating in 
internal loops.  In addition to probes for new physics, many of these modes
are also crucial to the full constraint of the ``Unitarity Triangle''.
As the definition implies, rare decays typically have branching fractions of 
less than $10^{-4}$.  The present data sample of roughly 88 million 
$B{\bar B}$ 
pairs allows for measurements or stringent limits on many such modes.

\subsection{Flavor and the Quark Sector of the Standard Model}

\begin{figure}[t]
\begin{center}
\begin{picture}(350,200)(-175,-130)
%\multiput(-175,-130)(0,10){21}{\line(1,0){350}}
%\multiput(-175,-130)(10,0){36}{\line(0,1){200}}
%
\thicklines
\put(-120,-80){\line(1,0){240}}
\put(-120,-80){\line(4,5){96}}
\put(+120,-80){\line(-6,5){144}}
\put(-65,-25){$V_{ud}V_{ub}^*$}
\put(+15,-25){$V_{td}V_{tb}^*$}
\put(-30,-75){$V_{cd}V_{cb}^*$}
\put(-25,+20){$\alpha$}
\put(+80,-70){$\beta$}
\put(-100,-70){$\gamma$}
\put(-130,-70){\vector(1,0){28}}
\put(-130,-70){\line(-1,-2){8}}
\put(-175,-130){\shortstack[l]{$B_S \rightarrow \rho K^0_S$ \\
                            $B^+ \rightarrow D^0_{CP} K^+$ \\
                            $B \rightarrow \pi\pi, K\pi$}}
\put(130,-70){\vector(-1,0){35}}
\put(130,-70){\line(1,-2){8}}
%**%\put(95,-120){\shortstack[l]{$B^0 \rightarrow%
\put(80,-120){\shortstack[l]{$B^0 \rightarrow%
                          (\psi^{(\prime)},\phi,\eta^\prime ) K^0_S$ \\
                          $B^0 \rightarrow D^{(*)+}  D^{(*)-}$}}
\put(  5, 28){\line(-1,3){8}}
\put(  5, 28){\vector(-3,-1){18}}
\put(-40, 55){$B \rightarrow \pi^+\pi^-, \rho^+\pi^- $}
\put(-30,-120){$b \rightarrow c\ell\nu$}
\put(-140,0){\shortstack[l]{$b \rightarrow u\ell\nu$\\
                            $B\rightarrow (\pi,\rho, \omega)\ell\nu$}}
\put(75,0){\shortstack[l]{$B^0_d$ Mixing \\
                         $B \rightarrow (\rho, \omega)\gamma$}}
\end{picture}
\end{center}
\caption{The Unitarity Triangle determined from the 
orthogonality of the first and third columns of the CKM matrix.
Also shown are B meson processes which yield information about 
each side and angle.} 
\label{fig:unit}
\end{figure}
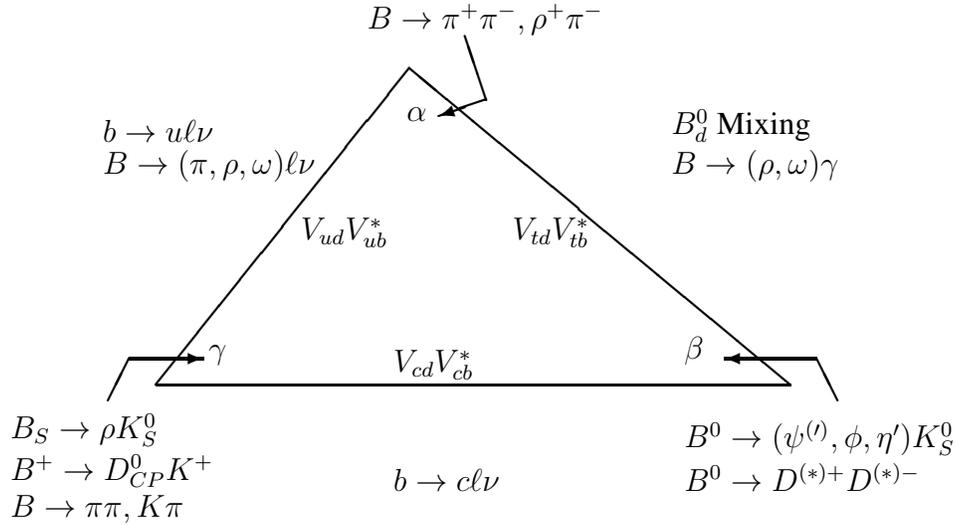

The complex CKM\cite{ckm} matrix describes the coupling of the charged weak
transition $q\rightarrow W^{*+}q^\prime$, which is proportional to 
$V^{*}_{qq^\prime}$.  The
non-diagonality of this matrix expresses the fact that the Weak isospin doublet
members $(b^\prime, s^\prime, d^\prime)$ are states of mixed flavor.  We can
thus view the CKM matrix as the transformation between the mass and
flavor eigenstates of the quarks
\begin{equation}
\left ( \begin{array}{c}
       d^\prime \\  s^\prime \\ b^\prime \end{array} \right)
       = \left ( \begin{array}{ccc}
       V_{ud} & V_{us} & V_{ub} \\
       V_{cd} & V_{cs} & V_{cb} \\
       V_{td} & V_{ts} & V_{tb} \end{array} \right)
       \left ( \begin{array}{c}
       d \\  s \\ b \end{array} \right). \label{eqn1}
\end{equation}
The unitarity condition implies that there are four 
free parameters in this matrix,
one of which is a phase.  It is through this phase that the Standard Model 
can accommodate CP violation.  In particular, the orthogonality requirement 
between the first and third columns requires
\begin{equation}
V_{ud}V_{ub}^*+V_{cd}V_{cb}^*+V_{td}V_{tb}^*=0 , \label{eqn2}
\end{equation}
which can be expressed geometrically as the so called ``Unitarity Triangle''
shown in Figure~\ref{fig:unit}.  Information about each side and angle is 
accessible through a variety of measurements in the B meson system.  The angles
are measured through time-dependent decay rate asymmetries, and the sides via 
direct or indirect measurements of the CKM matrix elements.  
Measurement of
all the components of the Unitarity Triangle over-constrains the triangle, and
thus provides a test of the Standard Model (SM).

\section{The \babar\ Detector}

A detailed description of the \babar\ detector can be found 
elsewhere\cite{babarnim}.  Charged particle momenta are measured
in a tracking system that consists of a 5-layer double-sided
silicon micro-strip vertex tracker (SVT) and a 40-layer drift chamber (DCH)
filled with an (80:20) mixture of helium and isobutane.  The
tracking volume is contained within the 1.5T magnetic
field of a superconducting solenoid.  
The combined track momentum resolution is $\sigma_{p_{T}}/p_{T}
=0.13\%\times p_{T} + 0.45\%$.  The primary charged hadron identifiation device
is a detector of internally reflected Cerenkov radiation (DIRC).  The typical
separation of kaons and pions due to their measured Cerenkov angle $\theta_C$
varies
from 8$\sigma$ at 2 GeV/c to 2.5$\sigma$ at 4 GeV/c, where $\sigma$ is the
average $\theta_C$ resolution.  Specific ionization energy loss (dE/dx) 
measurements in the DCH and SVT also contribute to charged hadron 
identification for particle momenta less than 0.7 GeV/c.
Photons are detected in an electromagnetic 
calorimeter (EMC) consisting of 6580 Thallium doped CsI crystals arranged in
barrel and forward end-cap sub-detectors.  The $\pi^0$ mass resolution in 
on average about 7 MeV/c$^2$.  Muons and long-lived neutral hadrons are 
detected within the instrumentation of the solenoid flux return (IFR) which 
consists of alternating layers of iron and resistive plate chambers.  

\section{Common Analysis Features}

\subsection{Data sample}

The analyses described here use all or part of a data sample consisting of 
approximately 88 million pairs of $\Upsilon (4S)\rightarrow B{\bar B}$ decays,
corresponding to a detector exposure of about 81 fb$^{-1}$.  
An additional sample of 
9.6 fb$^{-1}$ taken about 40 MeV below the peak of the $\Upsilon (4S)$ 
resonance
(``off-resonance'') is used by many analyses to study $e^+e^-\rightarrow 
q{\bar q}$ ``continuum'' backgrounds.

\subsection{B Meson Reconstruction}

B mesons produced from $\Upsilon (4S)$ decays are identified via their 
unique kinematics.  Because the mass of the $B$ meson pair is nearly that of
the $\Upsilon (4S)$, they are produced nearly at rest 
($p^*_B\approx 325$ MeV/c).  Use of the beam energy in constraining the 
kinematics serves to reduce the resolution of theses variables.

\begin{figure}[ht]
\begin{center}
%\leavevmode
\includegraphics[angle=-90,keepaspectratio=true,width=4.5in]{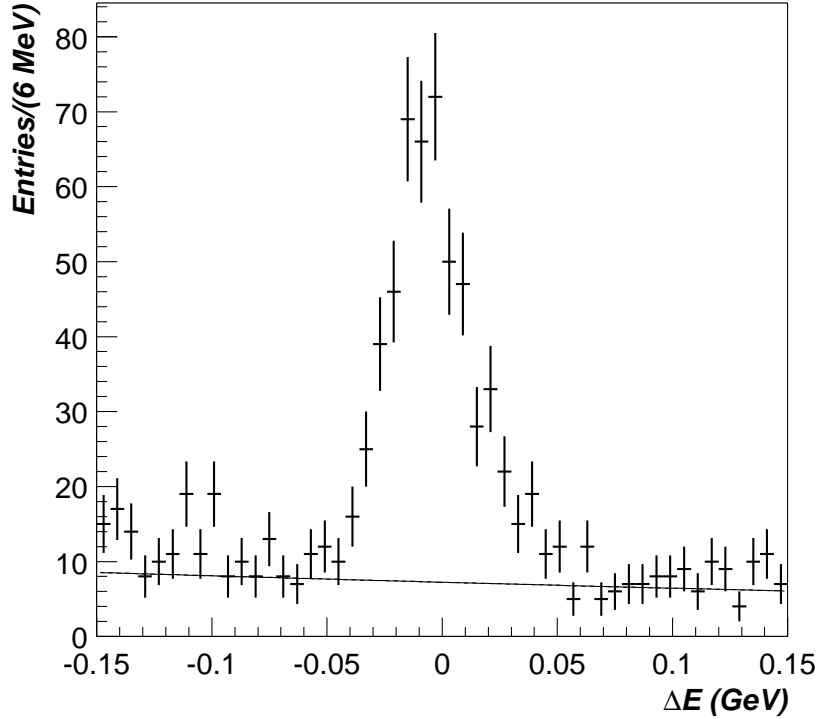}
\end{center}
\caption{$\Delta E$ distribution for charged $B$ decays to three charged kaons.
All candidate selection criteria have been applied except that on $\Delta E$.
The solid line shows the expected continuum background level.}
\label{fig:de}
\end{figure}

The conservation of energy can be expressed as:
\begin{equation}
\Delta E = E^*_B - E^*_{beam} , \label{eqn3}
\end{equation}
where $E^*_{beam}$ is the single beam energy in the center-of-mass (CM) 
frame.  
$E^*_B$ is the measured energy of the $B$ candidate in the CM.  
Correctly reconstructed $B$ candidates have $\Delta E$ distributed around 
zero with a resolution ranging from 15 to 80 MeV.  The energy resolution of 
the $B$ decay products dominates the resolution of this variable.  Continuum
background in this variable is well described by a monotonically decreasing 
low order polynomial.  
Figure~\ref{fig:de} shows the $\Delta E$ distribution for a typical rare mode 
after all selection criteria have been applied (except that on $\Delta E$).

\begin{figure}[ht]
\begin{center}
%\leavevmode
\includegraphics[angle=-90,keepaspectratio=true,width=4.5in]{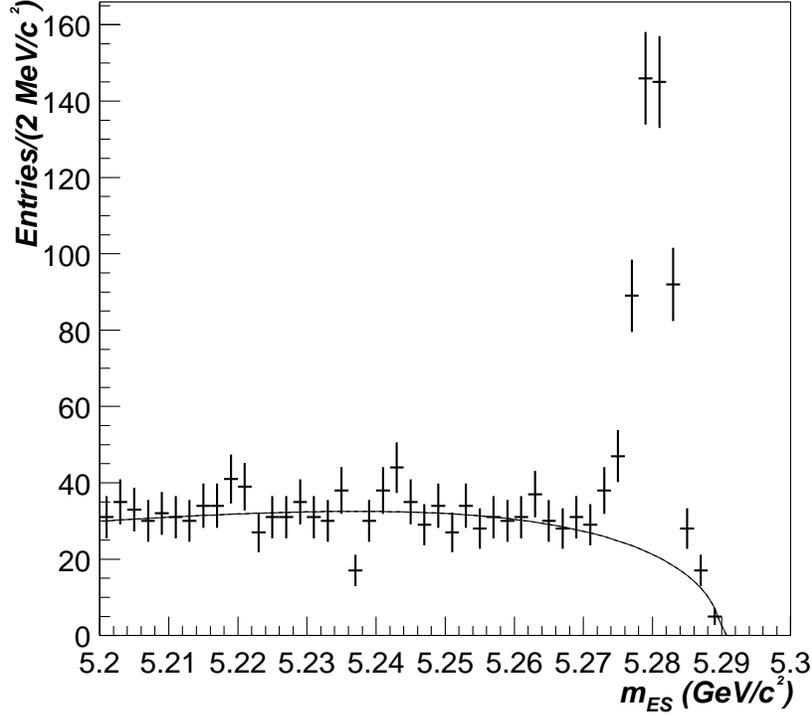}
\end{center}
\caption{$m_{ES}$ distribution for charged $B$ decays to three charged kaons.
All candidate selection criteria have been applied except that on $m_{ES}$.
The solid line shows the expected continuum background level.}
\label{fig:mes}
\end{figure}

We express momentum conservation as:
\begin{equation}
m_{ES} = \sqrt{E^{*2}_{beam} - {\vec p}^{*2}_B}  . \label{eqn4}
\end{equation}
Here $m_{ES}$ is the ``beam-energy substituted mass'', with ${\vec p}^*_B$
the $B$-candidate momentum in the CM.  Correctly reconstructed $B$ candidates
have $m_{ES}$ equal to the $B$ meson mass, with a resolution of about 
2.5-3.0 MeV/c$^2$, which is dominated by the beam energy spread.  The continuum
background shape in $m_{ES}$ is parameterized by a threshold 
function\cite{argus} with a fixed endpoint given by the average beam energy.
Figure~\ref{fig:mes} shows the $M_{ES}$ distribution for a typical rare mode
after all other selections have been applied.

In addition to the kinematics of the $B$ meson, signal events are selected
by making requirements on the decay products.  $B$ daughter resonances are 
required to have invariant masses within a restricted range typically 
determined by resolution and the need to leave sufficient sideband to determine
background levels.  Particle identification requirements are made to 
select some particles and veto sources of background.

\subsection{Background Suppression}

\begin{figure}[ht]
\begin{center}
%\leavevmode
\includegraphics[width=4.5in]{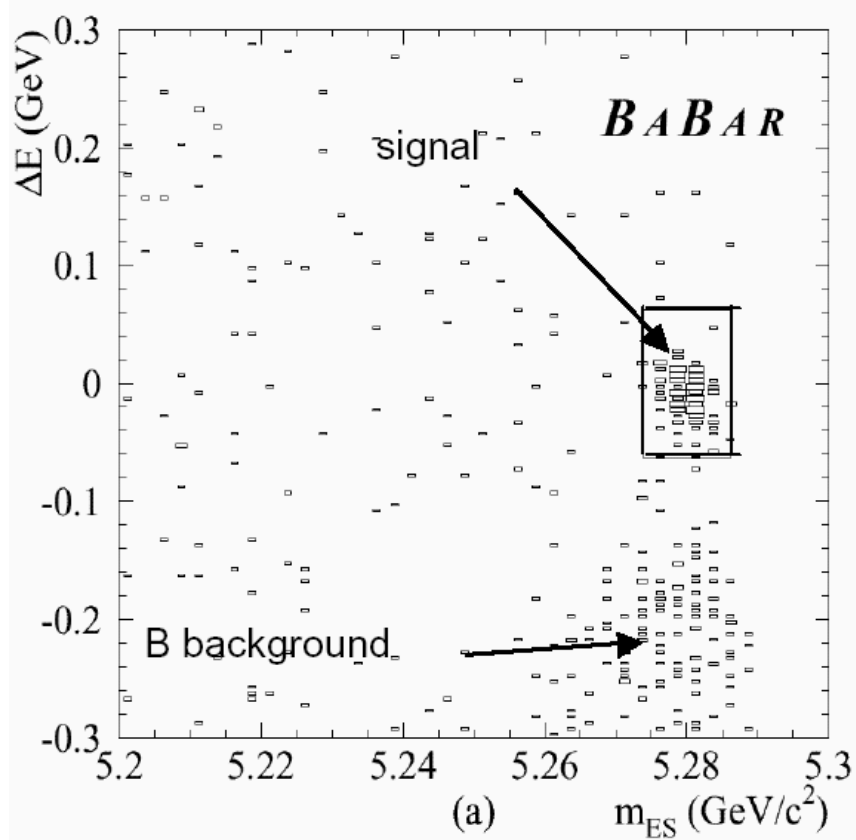}
\end{center}
\caption{Typical $m_{ES}$ vs $\Delta E$ distribution after event selection.
The signal populates the region around $m_{ES}=5.280$ and $\Delta E = 0$.
Continuum background populates the entire plane.  $B$
background populates the $\Delta E$ sideband.}
\label{fig:mesvsde}
\end{figure}

All rare analyses suffer from substantial backgrounds, and a variety of 
techniques are employed to reduce this to manageable levels.  In general, 
backgrounds from other $B$ decays are small.  Decays resulting from CKM
favored $b\rightarrow c$ transitions have heavier daughters and higher 
multiplicity final states than do CKM suppressed decays.  In order to wrongly
reconstruct such a decay as a rare signal, one must typically lose a particle 
from the true $B$ decay, resulting is a substantial shift in the candidate's
$\Delta E$.  The only exception to this is for rare modes with high final state
multiplicities, in which indications of some background have been observed.
While $B$ decays from other CKM suppressed transitions have similar kinematics
and multiplicities to that of a rare signal, such modes are rare themselves,
and require only limited suppression in most analyses.  Where $B$ backgrounds
are present, they typically populate the sidebands of the $\Delta E$ 
distribution, but have tails that reach into the signal region as illustrated
in Figure~\ref{fig:mesvsde}

For all the modes
discussed here, the primary background is due to random particle combinations
arising from continuum quark-antiquark production.  Although the probability 
for any given continuum event to satisfy a signal selection is quite small, 
the numbers favor the continuum.  The total production cross section
for light quarks (including charm) under the $\Upsilon (4S)$ is about 3.5 nb,
but only 1 nb for the $\Upsilon (4S)$ itself.  For a mode with an expected 
branching fraction of order 10$^{-6}$ this means that continuum events are 
produced at a rate well in excess of 10$^6$ times that of the signal.

In order to control continuum backgrounds, one typically exploits the 
fact that while the $B$ meson pairs are produced near threshold in 
$\Upsilon (4S)$ decays, the light and charm quark 
pairs which comprise the continuum
are produced with a great deal of excess energy.  The result is that 
for true $B$ meson decays, the decay products entering the detector are 
distributed isotropically in the CM, while the continuum background 
exhibits a ``jet-like'' topology, with a strong
correlation between the $B$ candidate decay and jet axes.

\begin{figure}[t]
\begin{center}
%\leavevmode
\includegraphics[angle=0,keepaspectratio=true,width=4.5in]{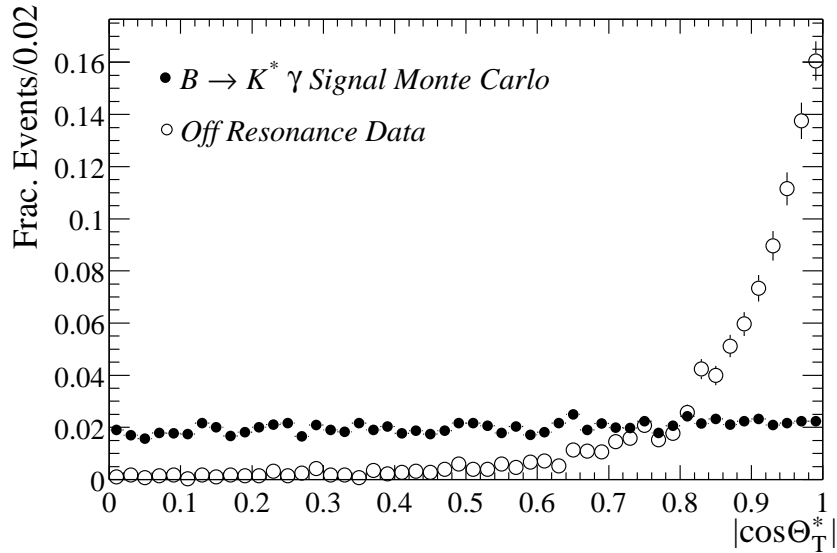}
\end{center}
\caption{Signal and background distributions of $|cos\theta_{Thrust}|$ for
a typical $B$ meson decay.  The signal distribution (solid histogram) is
uniform, reflection the random orientation between the $B$ candidate thrust 
axis and the thrust axis of the rest of the event.  The background 
(open points)
is strongly peaked at one due to the strong correlation between these two 
axes in fake $B$ candidates arising from random particle combinations in the
continuum.}
\label{fig:thrust}
\end{figure}

\begin{figure}[ht]
\begin{center}
%\leavevmode
\includegraphics[angle=-90,keepaspectratio=true,width=4.5in]{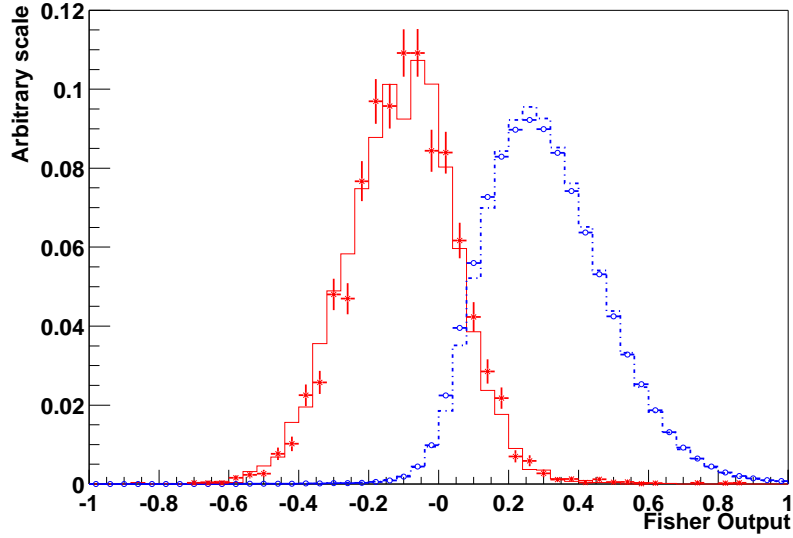}
\end{center}
\caption{Distribution of Fisher discriminant for data control mode (solid
points), control mode signal Monte Carlo (solid
histogram), continuum data (open points) and continuum Monte Carlo 
(dashed histogram) after a cut on 
$|cos\theta_{Thrust}|$.  The Fisher and Thrust angle are strongly correlated,
thus the separation will depend strongly on the thrust cut made.}
\label{fig:fisher}
\end{figure}

The first topological variable typically employed is the angle $\theta_T$ 
between the thrust axes of the $B$ candidate and the remaining particles in 
the event.  The sphericity axes may be used almost interchangeably.  The 
distribution of $|cos \theta_T|$ is nearly uniform for true $B$ mesons, but is 
strongly peaked near 1 for continuum background as is illustrated in 
Figure~\ref{fig:thrust}.  If additional background rejection is required, one 
may consider the remaining event shape information, such as the angles between 
the $B$ thrust and decay axes and the beam as well as the angular energy flow
in the event, and combine it into an optimized quantity using a neutral network
or a Fisher discriminant.  An example of the the separation power of a 
Fisher discriminant after a thrust cut has been made can be found in 
Figure~\ref{fig:fisher}.

\subsection{Signal Extraction}

All rare analyses at \babar\ are performed ``blind'', meaning that signal 
yields are hidden from the analyzer until the analysis has been peer 
reviewed and determined to be in a final form.  These steps are taken to 
avoid experimenter's bias.

There are two primary methods in which signal event yields are extracted; the 
event counting analysis, and the maximum likelihood fit.  Detection 
efficiencies determined from signal Monte Carlo simulations and data control
samples are used to convert yields into branching fraction measurements.  Equal
production of charged and neutral $B$'s from the $\Upsilon (4S)$ decay
are assumed throughout.

In the event 
counting analysis, a set of selection criteria are defined to select a signal 
region of the parameter space.  The criteria are optimized with respect to
expected signal and background yields to produce a measurement of the 
greatest possible statistical significance.  Systematic uncertainties may
figure into this process, but they are usually negligible for rare modes.  
The selection are applied to the data, and the population of signal region is 
counted.  An estimated background is subtracted to determine the signal event
yield.  The background yield is typically determined by measuring the density
of events in a sideband region and projecting that density into the signal
region.

In the maximum likelihood fit method signal yields are determined by an
unbinned extended maximum likelihood fit to a set of observables.  These 
typically include $m_{ES}$, $\Delta E$, event shape variables and, where 
appropriate, $B$ daughter resonance invariant
masses and particle identification information.  The 
probability 
${\cal P}_{i}({\vec x}_{j};{\vec \alpha}_{i})$ for a given hypothesis $i$ 
is the 
product of probability density functions (PDFs) for each of the variables
${\vec x}_{j}=(m_{ES}, \Delta E, {\rm Fisher}, ...)$ given 
the set of parameters
${\vec \alpha}_{i}$.  
The hypotheses $i$ are signal, continuum background and 
sometimes $B$ background for each final state in the fit.
The likelihood function is given by a product over all 
events $N$ and the signal and background components:
\begin{equation}
{\cal L} = \frac{e^{-\sum_{i}n_i}}{N!}\prod_{j=1}^{N}{\cal L}_{j},\;\;\; 
{\cal L}_{j} = \sum_{i}n_i{\cal P}_i({\vec x}_{j};{\vec \alpha}_{i}) .
\label{eqn5}
\end{equation}
The $n_i$ are the numbers of events for each hypothesis.  The values of the
yields (and any free parameters in the PDFs) are taken as those which maximize
 the likelihood function.  Unit change in $-2 ln {\cal L}$ defines the one 
standard deviation statistical uncertainties on the free parameters in the fit.
The statistical significance of the signal yield is determined from the
change in $-2 ln {\cal L}$ when the signal yield is forced to zero.  If
no statistically significant signal is found (more than 4 standard deviations),
a 90\% confidence level upper limit may be obtained by requiring:
\begin{equation}
\frac{\int_{0}^{n_{UL}} {\cal L}(n)dn}{\int_{0}^{\infty}{\cal L}(n)dn} = 0.9. 
\label{eqn6}
\end{equation}
Full and toy Monte Carlo simulations are used to verify that the fit is 
unbiased.

The accuracy with which the PDFs describe the data is of utmost importance in 
the likelihood fit.  Background PDFs are determined by fits to off-resonance and
sideband data.  Signal PDFs are determined primarily from signal Monte Carlo
simulations, but ultimately rely on data control samples to verify their 
validity.

\section{Electromagnetic Penguins}

\begin{figure}[t]
\begin{center}
%\leavevmode
\includegraphics[angle=0,keepaspectratio=true,width=3.5in]{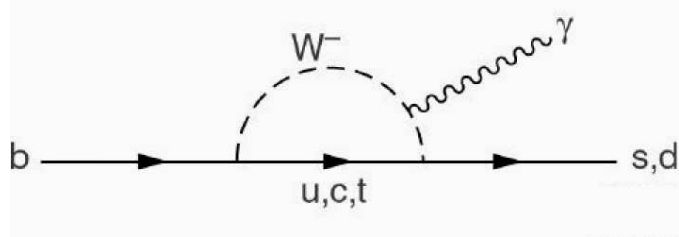}
\end{center}
\caption{Feynman diagram for an electromagnetic penguin amplitude.}
\label{fig:empenguin}
\end{figure}

Electromagnetic penguins consist of the class of amplitudes in which an 
external photon is emitted by one of the virtual particles 
participating in the loop through which the $b\rightarrow s(d)$ 
transition proceeds.
This is illustrated in Figure~\ref{fig:empenguin}.
Such diagrams are relatively clean from a theoretical perspective, and a 
variety of information can be gather from measurements of decays dominated
by these amplitudes.  The decay $B\rightarrow K^*\gamma$ was the first 
penguin to be 
observed\cite{kstar_cleo}.  Measurements of its branching fraction
provides a test of QCD, and direct CP violation in this mode would be an 
indication of new physics.  The decay rate ratio of $B\rightarrow\rho\gamma$ 
to $B\rightarrow K^*\gamma$ is sensitive to the ratio of 
$|\frac{V_{td}}{V_{ts}}|$.  The photon energy spectrum from measurements of 
$b\rightarrow s\gamma$ provides information on the mass and Fermi motion
of the $b$ quark within the $B$ meson.

In the analysis of these modes, in each case there is a requirement 
of a high energy isolated photon.  The calorimeter cluster is 
required to have a profile consistent with an electromagnetic shower, 
and the candidate photon must not be consistent with having originated from
a $\pi^0$ or $\eta$ decay.  Further details and results of each analysis are
presented below.

\subsection{Measurement of $B\rightarrow K^*\gamma$}

The analysis of $B\rightarrow K^*\gamma$\cite{babar_kstargamma} has been 
performed on a data sample corresponding to approximately 22 million 
$B{\bar B}$ pairs recorded in 1999-2000.
The $K^*\gamma$ final state is reconstructed in all four $K^*$ decay modes.
Stringent identification requirements are placed on charged kaons.  
Invariant mass
requirements are placed on both $K^*$ and $K^0_S$ candidates.  The $K^0_S$ is
also required to have a decay vertex displaced from the $e^+e^-$ interaction
point.  Since the $B$ meson is a pseudoscalar, angular momentum conservation
requires that the $K^*$ is polarized.  The absolute value of the cosine of 
the $K^*$ helicity angle is required to be less than 0.75.  
Continuum background is suppressed with cuts on the absolute values of the 
cosines of the thrust and $B$ flight angles, both of 0.80.

\begin{figure}[ht]
\begin{center}
%\leavevmode
\includegraphics[angle=0,keepaspectratio=true,totalheight=4.5in]{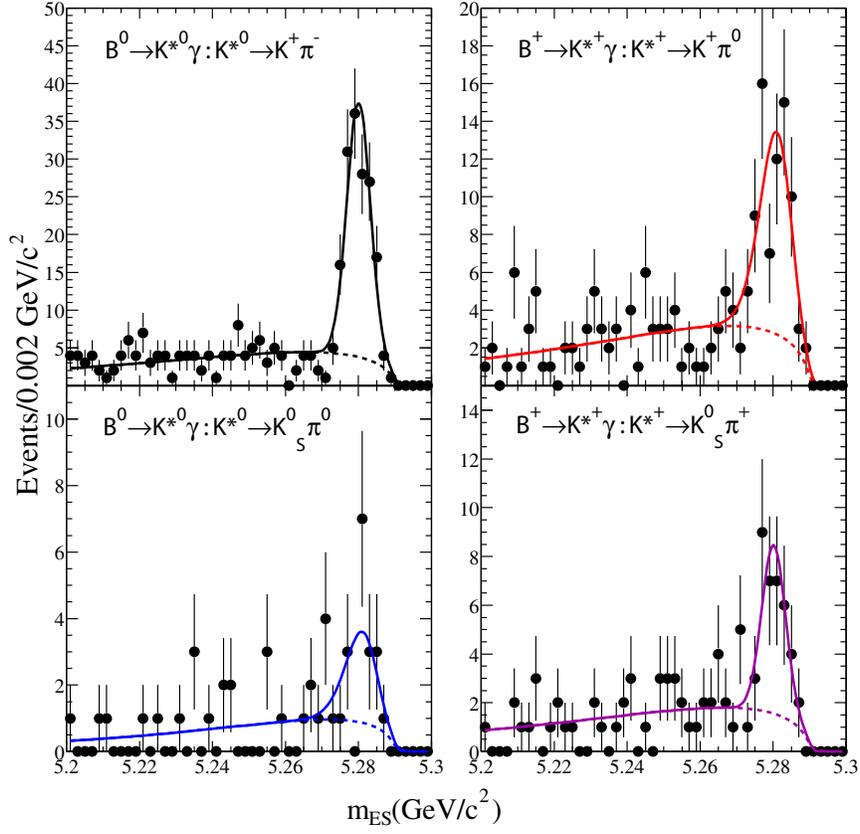}
\end{center}
\vspace{-0.25in}
\caption{The $m_{ES}$ distribution for $B\rightarrow K^*\gamma$ in each
of the $K^*$ decay modes.  All selection criteria have been applied.  
The solid curve is the combined signal and background
PDF shape, the dashed is the background only.}
\label{fig:kstar_mes}
\end{figure}

After cutting on $\Delta E$, the signal yield is determined from an 
unbinned maximum likelihood fit to the $m_{ES}$ distribution, shown for each
$K^*$ decay mode in Figure~\ref{fig:kstar_mes}.  Branching fraction and 
direct CP asymmetry results are shown in 
Table~\ref{tab:kstar_results}\cite{charge}.

\begin{table}[hb]
\begin{center}
%\hspace*{-1.5cm}
\begin{tabular}{|l|c|c|c|}
\hline
 & 
${\cal B}(B^0\rightarrow K^{*0}\gamma)$ & 
${\cal B}(B^+\rightarrow K^{*+}\gamma)$ &
$A_{CP}$ \\
\hline
\hline
Theory\cite{kstar_theory1,kstar_theory2,kstar_theory3} & 
$7.5 \pm 3.0$ &
$7.5 \pm 3.0$ &
$|A_{CP}|<0.005$ \\
\hline
\babar &
$4.23 \pm 0.40 \pm 0.22$ &
$3.83 \pm 0.62 \pm 0.22$ &
$-0.17 < A_{CP} < 0.08$ \\
& & & @90\% CL \\
\hline
\end{tabular}
\end{center}
\caption{Results of the branching fraction and direct asymmetry analysis
of $B\rightarrow K^{*}\gamma$.  In each result, the first uncertainty is 
statistical, the second systematic.  Branching fractions are in units of 
$10^{-5}$.}
\label{tab:kstar_results}
\end{table}

\subsection{Search for $B\rightarrow\rho\gamma$ and $B\rightarrow\omega\gamma$}

The analysis of the $\rho\gamma$ and $\omega\gamma$ final 
states\cite{babar_rhogamma} is 
significantly more challenging than that of $K^*\gamma$.  The predicted 
branching fractions are about 50 times smaller than for $K^*\gamma$.  Both
the $\rho$ and the $\omega$ have significantly more background under
their peaks than does the $K^*$, and the rho is much broader.  In addition 
to continuum background, these modes also potentially suffer 
cross-feed background from 
$K^*\gamma$, other $b\rightarrow s\gamma$ processes and 
from $B\rightarrow\rho\pi^0$.

A neutral network containing information from event shape, $\Delta t$ and 
flavor tagging is used to control continuum background.  $K^*\gamma$ 
feed-across is vetoed using particle identification.  After these selection
criteria are applied, the signal yield for
each final state is extracted using a unbinned maximum likelihood fit to 
$m_{ES}$, $\Delta E$, and the $\rho / \omega$ invariant mass.  Studies of 
generic $B{\bar B}$ Monte Carlo show that the expected $B$ background is 
quite small, so the fit includes components only for signal and continuum
background.  Background from $B$ decays is considered as a systematic 
uncertainty.

The results of this analysis applied to a sample of 84 million $B{\bar B}$ 
pairs can be found in table~\ref{tab:rhogamma_results}.  If isospin symmetry
is assumed, all three modes can be combined to produce an upper limit of
${\cal B}(B\rightarrow\rho\gamma)<1.9\times 10^{-6}$ @ 90\% CL.  This result
can be used to place on upper limit on CKM parameters 
$|\frac{V_{td}}{V_{ts}}| < 0.036$ @ 90\% CL.  A discussion of theoretical
errors can be found in Ali and Parkhomonko\cite{kstar_theory3}.

\begin{table}[h]
\begin{center}
%\hspace*{-1.5cm}
\begin{tabular}{|l|c|c|c|}
\hline
 & 
${\cal B}(B^0\rightarrow \rho^{0}\gamma)$ & 
${\cal B}(B^+\rightarrow \rho^{+}\gamma)$ &
${\cal B}(B^0\rightarrow \omega\gamma)$ \\
\hline
\hline
Theory\cite{kstar_theory3} & 
$0.5-0.75$ &
$0.8-1.5$  &
$0.5-0.75$ \\
\hline
\babar &
$< 1.4$ &
$< 2.3$ &
$< 1.2$ \\
\hline
\end{tabular}
\end{center}
\vspace{-0.4cm}
\caption{Results of the branching fraction analysis
of $B\rightarrow \rho\gamma$ and $\omega\gamma$.  
Branching ratios are in units of $10^{-6}$.  Upper limits are at 90\% CL.}
\label{tab:rhogamma_results}
\end{table}

\subsection{Semi-Inclusive Measurement of $b\rightarrow s\gamma$}

This analysis\cite{babar_semibsg} of 22 million $B{\bar B}$ pairs 
is a study of a collection of
exclusive final states with a kaon plus up to four pions, no more than one of 
which may be neutral.  
Because $b\rightarrow s\gamma$ is a two-body decay
process, the photon energy $E_{\gamma}$ in the $B$ rest frame is related to
the recoil hadronic mass, $M_{Had}$:
\begin{equation}
E_{\gamma} = \frac{M_B^2-M_{Had}^2}{2M_B}.
\label{eqn7}
\end{equation}
Fits to the measured spectra of both of these quantities can be used to 
determine the total branching ratio for 
$B\rightarrow X_s\gamma$\cite{bsg_theory1}.  In addition to constraining 
new physics contributions to the underlying amplitude, parameters associated 
with heavy quark effective theory (HQET) are also extracted in the analysis.
These parameters are critical to reducing theory errors in the extraction of
$V_{ub}$ and $V_{cb}$.

\begin{figure}[ht]
\begin{center}
%\leavevmode
\psfiletwo{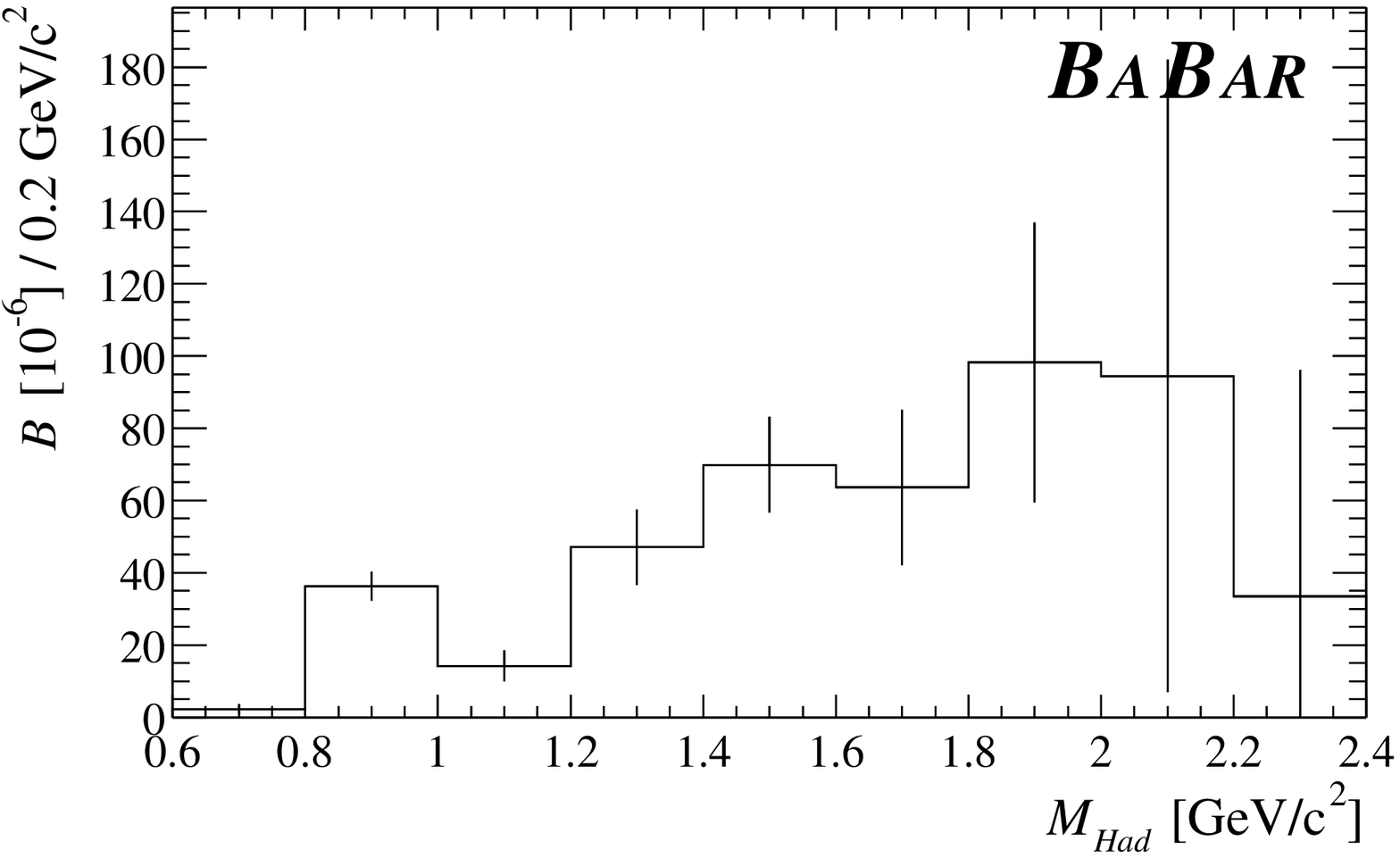}{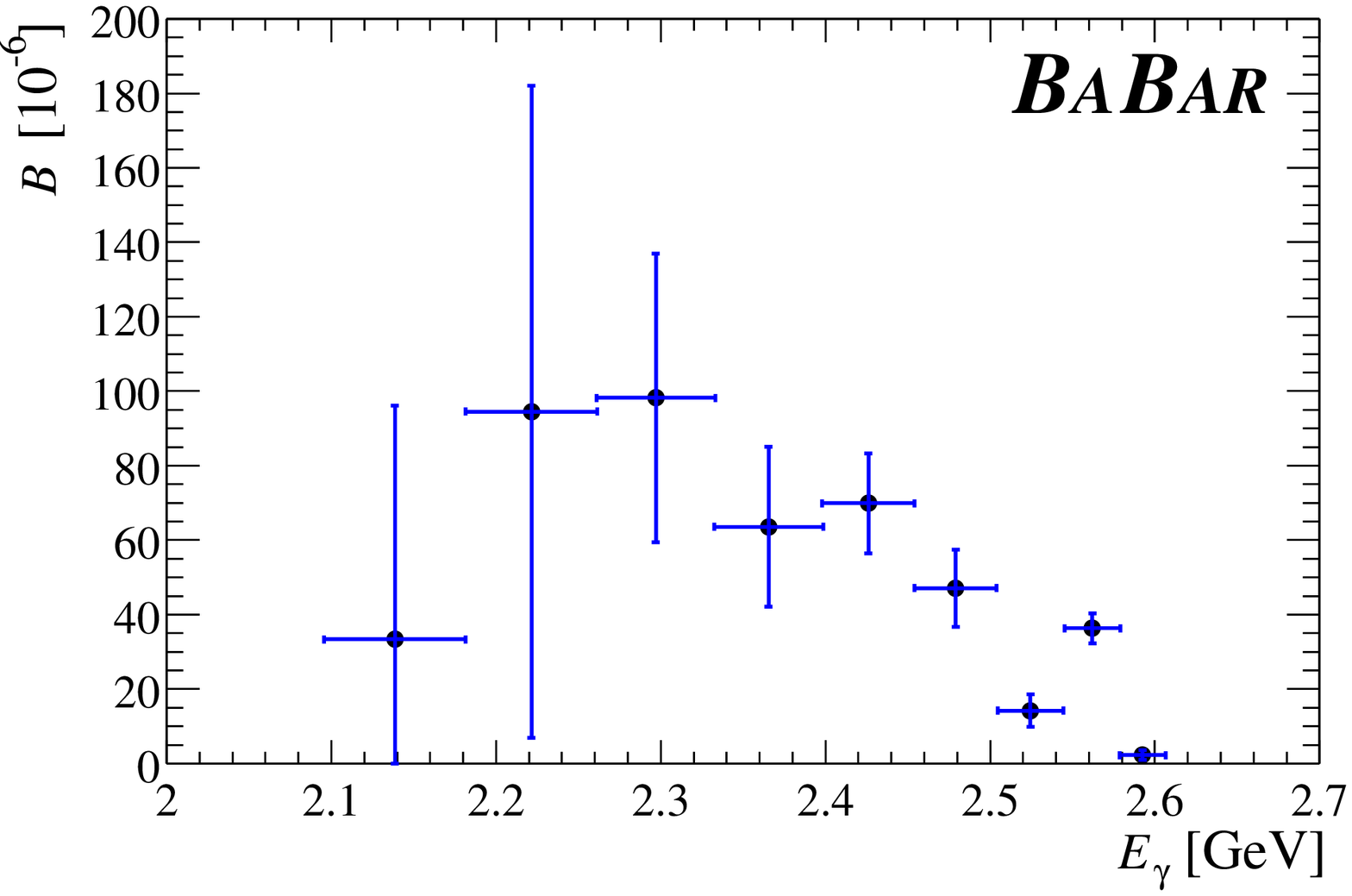}{1.0}%
\end{center}
\caption{Hadronic recoil mass and photon energy spectra for the semi-inclusive
analysis of $b\rightarrow s\gamma$.  Errors are statistical only.}
\label{fig:semi_bsgspectra}
\end{figure}

Measured branching fractions as a function of $M_{Had}$ and $E_{\gamma}$ can 
be found in Figure~\ref{fig:semi_bsgspectra}.  Analysis of these spectra 
yield results:
\begin{eqnarray}
{\bar \Lambda} & = & 0.37 \pm 0.09 (stat) \pm 0.07 (syst) \pm 0.10 (model) 
\;{\rm GeV/c^2} \nonumber \\
m_b & = & 4.79 \pm 0.08 (stat) \pm 0.10 (syst) \pm 0.10 (model) 
\;{\rm GeV/c^2} \nonumber \\
\lambda_1 & = & -0.24^{+0.03}_{-0.04}\; (stat) 
\pm 0.02 (syst)\; ^{+0.15}_{-0.21} 
(model) 
\;{\rm GeV/c^2} \nonumber \\
{\cal B}(b\rightarrow s\gamma) & = & 4.3 \pm 0.5 (stat) \pm 0.8 (syst) 
\pm 1.3 (model) \times 10^{-4}.
\label{eqn11}
\end{eqnarray}

\subsection{Fully-Inclusive Measurement of $b\rightarrow s\gamma$}

Much of the uncertainty in the semi-inclusive $b\rightarrow s\gamma$ analysis
arises from theoretical errors.  HQET implies a duality between the quark and
hadron level of an interaction, which implies that parton level rate for 
$b\rightarrow s\gamma$ is the same as the inclusive rate for 
$B\rightarrow X_s\gamma$.  These two issues motivate the fully inclusive 
analysis technique.  

This analysis\cite{babar_fullbsg} 
is performed on a sample of 60 million $B{\bar B}$ pairs.  Photons in the 
range $1.5 < E^*_\gamma < 3.5$ GeV are analyzed.  
These photons are required to meet 
the selection criteria described above.  To suppress continuum background
the event is required to have a lepton flavor tag which strongly selects
true $B{\bar B}$ decays.  Such a selection induces no model dependency into the
analysis as it only applies to the ``other'' $B$ in the decay. 
In addition to particle identification criteria,
fake leptons are further rejected by requiring large missing energy in the 
event, which is normally associated with semi-leptonic $b\rightarrow c$ 
transitions.  To obtain additional discrimination, the angular separation
between the lepton and the photon is required not to be small.  Event topology
in the form of the Fox-Wolfram moments in also employed to reduce background
from the continuum.  Backgrounds are estimated using off-resonance data and 
$B{\bar B}$ Monte Carlo.

\begin{figure}[ht]
\begin{center}
%\leavevmode
\includegraphics[angle=0,keepaspectratio=true,totalheight=3.25in]
{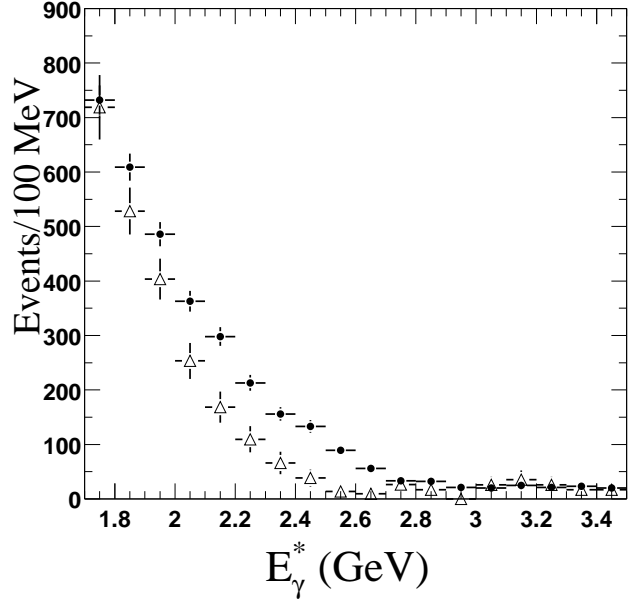}
\end{center}
\vspace{-0.3in}
\caption{The $E^*_\gamma$ distribution of on-resonance data (solid points)
and background expectations for the fully inclusive analysis
of $B\rightarrow X_s\gamma$.  Errors are statistical only.}
\label{fig:inclbsg_egam}
\end{figure}

The $E_\gamma$ spectra for on-resonance data and the predicted background are
shown in Figure~\ref{fig:inclbsg_egam}.  
The photon energy range
$2.1 < E^*_\gamma < 2.7$ GeV is considered to reduce model dependencies.  The 
branching fraction for $B\rightarrow X_s\gamma$ is measured in this region and
then extrapolated to the full spectrum:
\begin{equation}
{\cal B}(B\rightarrow X_s\gamma) = 3.88 \pm 0.36 (stat) \pm 0.37 (syst) 
^{+0.43}_{-0.23} (model) \times 10^{-4}. 
\label{eqn12}
\end{equation}

\section{Electroweak Penguins}

As the name suggests, Electroweak penguins are amplitudes that proceed via 
loops involving photons, $W$ or $Z$ bosons.  Such processes are strongly 
suppressed in the Standard Model, and as such, are excellent windows onto
potential new physics.  Well controlled theoretical uncertainties aid in this 
sensitivity.  We discuss the analysis of four such final states 
here: $K^{(*)}\ell^+\ell^-$\cite{babar_kll}, 
$\ell^+\ell^-$\cite{babar_ll},
$K\nu{\bar \nu}$\cite{babar_knunu}, 
and $\gamma\gamma$\cite{babar_2gamma}.

\subsection{Measurement of $B\rightarrow K^{(*)} \ell^+\ell^-$}

\begin{figure}[ht]
\begin{center}
%\leavevmode
\includegraphics[angle=0,keepaspectratio=true,width=4.5in]
{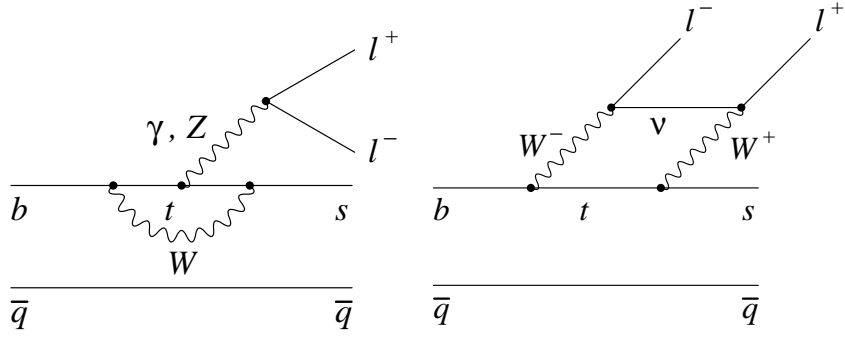}
\end{center}
%\vspace{-0.3in}
\caption{Leading Electroweak penguin and box diagrams for the decay
$B\rightarrow K^{(*)} \ell^+\ell^-$.}
\label{fig:kll_diagram}
\end{figure}

The flavor-changing neutral current decays $B\rightarrow K \ell^+\ell^-$
and $B\rightarrow K^* \ell^+\ell^-$ have predicted branching fractions 
on the order $10^{-6}-10^{-7}$\cite{kll_theory1}.
The leading diagrams for this decay are electroweak penguin and box diagrams, 
and can be found in Figure~\ref{fig:kll_diagram}.

The decay rate for $B\rightarrow K^{(*)} \ell^+\ell^-$ is rather sensitive to
the presence of new physics.  In particular, certain extensions to the SM can 
vary the rate by more than a factor of two.  In addition to the decay rate,
kinematic distributions accessible with higher statistics, such as the boson
$q^2$ distribution ($m_{\ell\ell}^2$) and the 
forward-backward asymmetry in the $K^*$ channel are of considerable 
interest as they are also quite
sensitive to non-SM physics, and are less model dependent than the overall
rate.  

The experimental challenge in this analysis is to control the various sources 
of background.  Background from $B\rightarrow$ charmonium decays which have the
same final state particles are control by vetoing regions in the 
$\Delta E$ vs $m_{\ell\ell}$ plane.  Continuum background is reduced
using a Fisher discriminant which in addition to event shape information 
includes information on the $K\ell$ invariant mass, which serves to veto 
$D\rightarrow K\ell\nu$.  Combinatorics from semi-leptonic $B$ decays
are rejected using a B-likelihood built from the missing energy in the event,
vertex information, and the $B$ production angle.  Finally, peaking backgrounds
from particle mis-identification are reduced by vetoing the $K^{(*)}\pi$ mass
in the region of the $D$ mass.

\begin{figure}[ht]
\begin{center}
%\leavevmode
\includegraphics[angle=0,keepaspectratio=true,height=4.5in]
{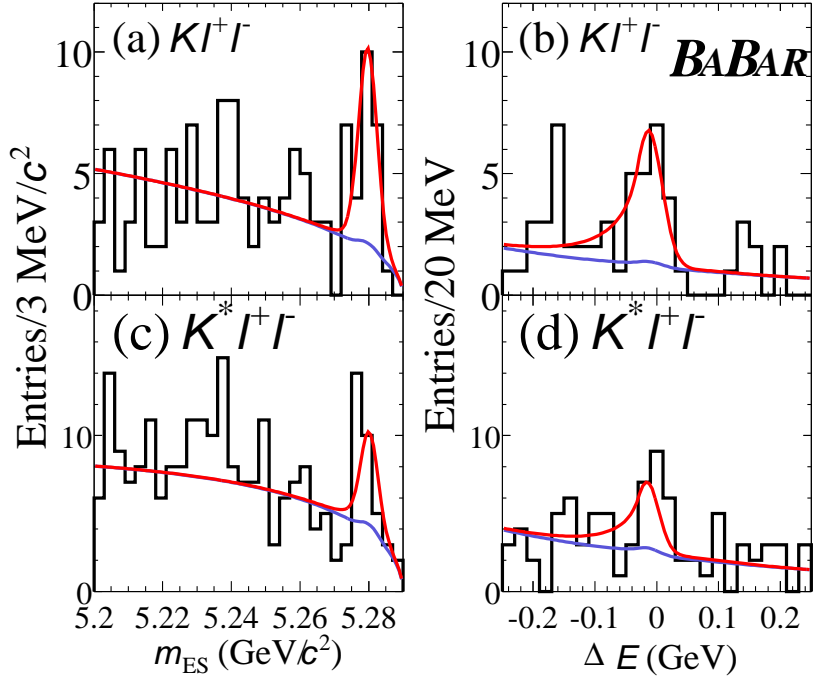}
\end{center}
\vspace{-0.65in}
\caption{Projections of $m_{ES}$ and $\Delta E$ for the combined
$K^{(*)}\ell^+\ell^-$ final states.  The solid line is the combined signal
plus background PDF, the histogram are the data.}
\label{fig:kll_proj}
\end{figure}

After background rejection and particle identification criteria are applied, 
the signal is extracted with a likelihood fit to $m_{ES}$ and $\Delta E$.
The results of this analysis on a sample of 88.4M $B{\bar B}$ pairs are:
\begin{equation}
{\cal B}(B\rightarrow K\ell^+\ell^-) = (0.78^{+0.24}_{-0.20} (stat) 
^{+0.11}_{-0.18} (syst) ) \times 10^{-6} 
\label{eqn13}
\end{equation}
with a significance (including systematics) of $4.4\sigma$, and 
\begin{equation}
{\cal B}(B\rightarrow K^*\ell^+\ell^-) = (1.68^{+0.68}_{-0.58} (stat) 
\pm 0.28 (syst) ) \times 10^{-6}
\label{eqn14}
\end{equation}
with a significance of $2.8\sigma$.  Since the $K^*$ result
is not significant, we report a 90\% CL upper limit:
\begin{equation}
{\cal B}(B\rightarrow K^*\ell^+\ell^-) < 3.0 \times 10^{-6}. 
\label{eqn15}
\end{equation}
Combined projections of $m_{ES}$ and $\Delta E$ are shown in 
Figure~\ref{fig:kll_proj}.
%\vspace{-0.4in}

\subsection{Search for $B^0\rightarrow \ell^+\ell^-$}

\begin{figure}[ht]
\begin{center}
%\leavevmode
\includegraphics[angle=0,keepaspectratio=true,width=4.5in]
{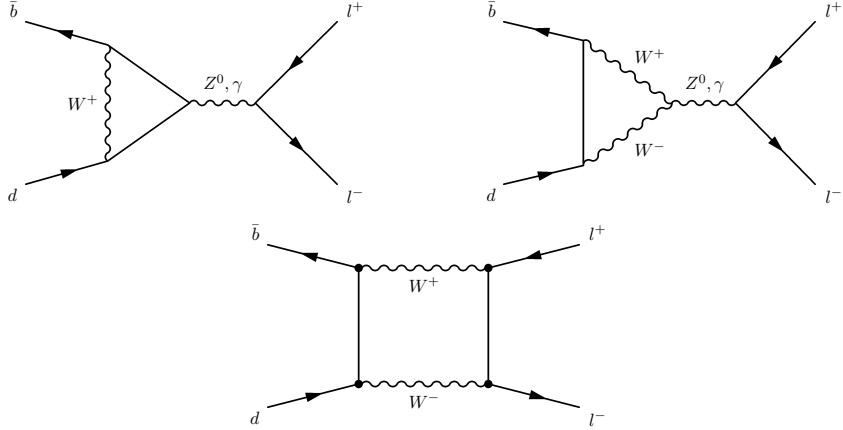}
\end{center}
%\vspace{-0.3in}
\caption{Leading diagrams for the decay
$B\rightarrow \ell^+\ell^-$.}
\label{fig:ll_diagrams}
\end{figure}

The decay of a $B$ meson to a pair of leptons is highly suppressed within the
SM by factors resulting from CKM, internal quark annihilation and helicity.
Leading diagrams are shown in Figure~\ref{fig:ll_diagrams}.  Within the SM,
predicted branching fractions are $1.9\times 10^{-15}$ and $8.0\times 10^{-11}$
for the $e^+e^-$ and $\mu^+\mu^-$ channels respectively\cite{ll_theory1}.  
The $e\mu$ channel is forbidden by lepton number conservation.  New physics can
significantly alter these predictions\cite{ll_theory2}.

In this analysis, the primary sources of background are from real lepton
production from continuum $c{\bar c}$ decays, pions which are mis-identified as
muons, and two photon processes.  Continuum background is suppressed using the 
thrust magnitude, and the angle between the thrust axes of the $B$ candidate
and the rest of the event.  A track multiplicity cut serves to reject 
two photon processes.  The signal is selected by requiring two high momentum
leptons of opposite charge and good vertex information.  Particle 
identification requirements for both leptons are made.  The signal yield is
determined by counting events in a signal region of $m_{ES}$ and $\Delta E$,
and subtracting an estimated background determined from the scaled population
of the $m_{ES}$ vs $\Delta E$ plane.  The results of this analysis
applied to a sample of approximately 60 million $B{\bar B}$ pairs are 
presented in Table~\ref{tab:ll_results}.

\begin{table}[ht]
\begin{center}
%\hspace*{-1.5cm}
\begin{tabular}{|l|c|c|c|c|}
\hline
 & $N_{GSB}$ & $N_{SigBox}$ & $N_{BG}$ & 90\% CL Upper Limit \\
\hline
\hline
${\cal B}(B^0\rightarrow e^+e^-)$ & 25 & 1 & $0.60 \pm 0.24$ 
& $3.3 \times 10^{-7}$ \\
\hline
${\cal B}(B^0\rightarrow \mu^+\mu^-)$ & 26 & 0 & $0.49 \pm 0.19$ 
& $2.7\times 10^{-7}$ \\
\hline
${\cal B}(B^0\rightarrow \mu^+\mu^-)$ & 26 & 0 & $0.49 \pm 0.19$ 
& $2.7\times 10^{-7}$ \\
\hline
\end{tabular}
\end{center}
\caption{Results of the search for
$B\rightarrow\ell^+\ell^-$.  $N_{GSB}$ is the population of the 
$m_{ES}$ vs $\Delta E$ sideband and $N_{SigBox}$ the population of the signal
region after all selection criteria have been applied.  $N_{BG}$ 
is the expected
background in the signal region based on the sideband population.}
\label{tab:ll_results}
\end{table}

\subsection{Search for $B^+\rightarrow K^+ \nu{\bar \nu}$}

Within the SM, the decay $b\rightarrow s \nu{\bar \nu}$ is a pure 
electroweak flavor changing neutral current.  The final state is nearly free
of strong interaction uncertainties, and hence the theoretical errors 
associated with this decay are small.  While the inclusive analysis is not
currently feasible, it is possible to search for the exclusive decay 
$B^+\rightarrow K^+ \nu{\bar \nu}$.  Summing over all neutrino species,
the SM prediction for this branching fraction\cite{babar_physbook} is
\begin{equation}
{\cal B}(B^+\rightarrow K^+\nu{\bar \nu}) = 3.8^{+1.2}_{-0.6} \times 10^{-6}. 
\label{eqn16}
\end{equation}

The presence of two neutrinos in the final state makes this analysis difficult,
as there are no kinematic constraints which may be applied to the signal $B$.
Instead, the strategy is to fully reconstruct the other $B$ from the 
$\Upsilon (4S)$ decay, and compare the remaining particles in the event with
the signature expected from the signal.  The ``tag'' $B$ is required to be
fully reconstructed as either $B^-\rightarrow D^0\ell^-{\bar \nu}$ or
$B^-\rightarrow D^{*0}\ell^-{\bar \nu}$.  The $D^0$ is reconstructed in the 
$K^-\pi^+$, $K^-\pi^+\pi^-\pi^+$ and $K^-\pi^+\pi^0$ modes, which results in a 
total of about 0.5\% of all charged $B$s being reconstructed as tags.  To
select signal events, a high momentum charged kaon is required in the 
recoil of the tagged $B$.  Additional requirements are made on the neutral 
energy in the recoil and on the angle between the kaon and the tag side lepton.

\begin{figure}[ht]
\begin{center}
%\leavevmode
\includegraphics[angle=0,keepaspectratio=true,width=4in]
{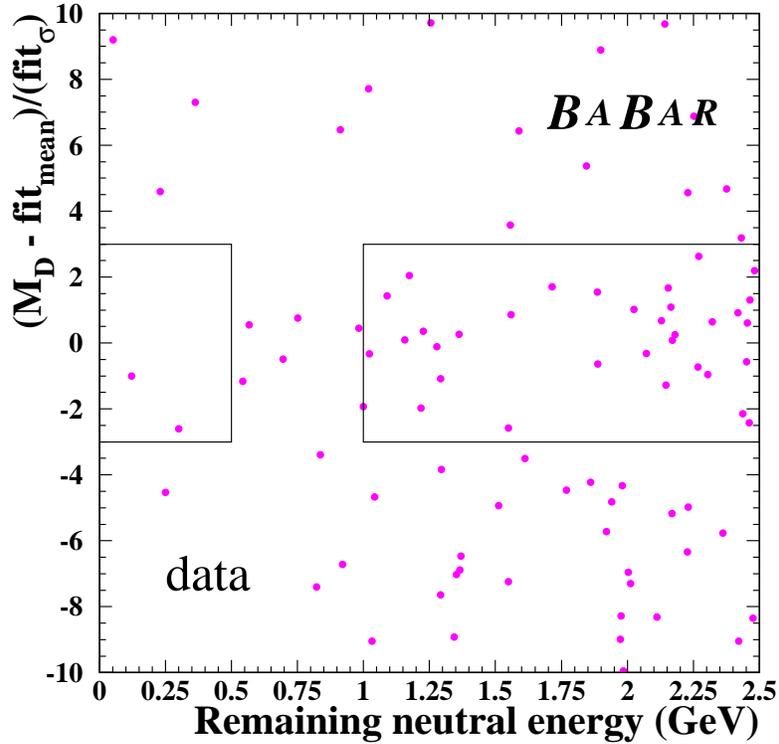}
\end{center}
%\vspace{-0.3in}
\caption{Distribution of $K^+\nu{\bar \nu}$ candidates in the plane
defined by the electromagnetic energy not from the tag $B$, $E_{Left}$, 
and the resolution scaled difference between the measured and 
mean measured $D$ mass.  The signal region is the bounded area at the 
left of the plot.}
\label{fig:knunu_data}
\end{figure}

Events are counted in a signal region in the 
plane defined by the electromagnetic energy in the tag $B$ recoil, and 
the difference between the reconstructed and
mean fitted $D$ mass, scaled by the fitted $D$ mass resolution.  The 
expected background, determined by scaling the sideband 
population into the signal region, is subtracted from the signal region 
population to determine the signal yield.  This is illustrated in 
Figure~\ref{fig:knunu_data}.  In a sample of 60 million $B{\bar B}$ pairs,
the expected background in the signal region is 2.2 events, and there are two 
events observed.  The 90\% confidence level upper limit 
on the branching fraction, including systematics, is
\begin{equation}
{\cal B}(B^+\rightarrow K^+\nu{\bar \nu}) < 9.4 \times 10^{-5}. 
\label{eqn17}
\end{equation}

\subsection{Search for $B^0\rightarrow \gamma\gamma$}

The decay $B^0\rightarrow \gamma\gamma$ is an example of electroweak 
annihilation.  The SM expectation for this decay is small, with predictions
ranging from 0.1 to $2.3\times 10^{-8}$\cite{2gamma_theory1}.  As with the 
other modes discussed in this section, physics beyond the SM can result in 
significant enhancements to this rate\cite{2gamma_theory2}.

In this analysis, selection criteria are placed on the ratio of the 2nd to 0th
Fox-Wolfram moments, the cosine of the angle between one of the photons 
(chosen at random) and the thrust axis of the rest of the event, and the $B$
production angle to suppress continuum background.  Selected photons are 
required not to be consistent with having come from a $\pi^0$ or $\eta$ decay.
The signal yield is determined by counting events in a signal region of the 
plane defined by $m_{ES}$ and $\Delta E$, and subtracting the expected 
background determined by scaling the the sideband population into the signal
region.  The result for a sample of 22 million $B{\bar B}$ pairs is
\begin{equation}
{\cal B}(B^0\rightarrow\gamma\gamma) < 1.7 \times 10^{-6},
\label{eqn18}
\end{equation}
at the 90\% confidence level, including systematic uncertainties.

\section{Gluonic Penguins (Charmless Hadronic B Decays)}

Charmless hadronic $B$ decays proceed through a combination of CKM suppressed
tree ($b\rightarrow u$) and gluonic penguin ($b\rightarrow d,s$) amplitudes.
There are about 70 possible combinations of two-body decays in the lowest 
pseudoscalar and vector nonets.  These may be further broken into two groups;
two-body decays in which both $B$ daughters are kaons or pions, and 
quasi-two-body decays in which at least one of the $B$ daughters is a 
short-lived resonance.  The two-body modes can be analyzed for information 
on the CP phases $\alpha$ and $\gamma$, and have been found to have significant
penguin 
contributions in addition to CKM allowed tree amplitudes.  Several of the 
quasi-two-body modes are sensitive to the CP phase $\beta$.  
In addition to yielding
information about the Unitarity Triangle, decays in which penguin amplitudes
are dominant are sensitive to new physics.  Our study of three-body $B$ decays
has thus far been limited to combinations of three charged kaons or pions.

All of these modes share some common features.  The primary source of 
background is random particle combinations in the continuum, although modes
with large final state multiplicities or significant neutral energy may suffer
from non-negligible $B{\bar B}$ backgrounds.  All the final states are
ultimately composed of high momentum kaons and pions, so the ability to 
distinguish been these particles at high momenta is crucial.

\subsection{Two-Body Decays}

Two body $B$ decays to kaons and pions are sensitive to the angle $\alpha$ 
of the Unitarity Triangle through the time-dependent CP violating asymmetry 
in the decay $B\rightarrow\pi^+\pi^-$ and 
to the the angle $\gamma$ through branching fractions and direct CP-violating
asymmetries of decays to various $\pi\pi$ and $K\pi$ final states.  Because 
there are substantial penguin amplitudes which contribute to the $\pi^+\pi^-$ 
final state in addition to the tree amplitude, the time-dependent asymmetry in 
that mode does not directly measure $\alpha$.  An isospin analysis of the rates
for all the $B\rightarrow\pi\pi$ decays is required to fully unfold the 
effects of 
the penguin contributions and determine the relationship between what is 
measured from the $\pi^+\pi^-$ analysis ($\alpha_{eff}$) and $\alpha$.  
Interference between 
penguin and tree amplitudes may also lead to substantial direct 
(time-independent) CP asymmetries in the $K\pi$ final states.

In each mode, the signal is extracted using an unbinned extended maximum 
likelihood fit, using the $m_{ES}$, $\Delta E$, a Fisher discriminant, and
where appropriate, Cerenkov angle residuals.  Groups of related decays are fit
simultaneously.  For example, the  $\pi^+\pi^-$, $K^+\pi^-$ and 
$K^+K^-$ yields are determined from a single fit.  In these cases, $\Delta E$ 
and the Cerenkov angle residuals separate the signal modes from each other.
Branching fraction results for all two-body modes based on a sample of
88 million $B{\bar B}$ pairs can be found in 
Table~\ref{tab:2body}.

\begin{table}[ht]
\begin{center}
%\hspace*{-1.5cm}
\begin{tabular}{|c|c|c|c|}
\hline
Decay & $N_{Signal}$ & ${\cal B}\times 10^{-6}$ & $A_{CP}$\\
\hline
\hline
$B^0\rightarrow\pi^+\pi^-$        & $157 \pm 19 $ & $4.7 \pm 0.6 \pm 0.2$ & \\
\hline
$B^0\rightarrow K^{\pm}\pi^{\mp}$ & $589 \pm 30 $ & $17.9 \pm 0.9 \pm 0.7$ & 
$-0.102 \pm 0.050 \pm 0.016$\\
\hline
$B^0\rightarrow K^+K^-$           & $1 \pm 8 $ & $<0.6$ & \\
\hline
$B^+\rightarrow\pi^+\pi^0$        & $125 \pm 22 $ & $5.5 \pm 1.0 \pm 0.6$ & 
$-0.03 \pm 0.18 \pm 0.02$ \\
\hline
$B^+\rightarrow K^+\pi^0$        & $239 \pm 22 $ & $12.8 \pm 1.2 \pm 1.0$ & 
$-0.09 \pm 0.09 \pm 0.01$ \\
\hline
$B^+\rightarrow K^0\pi^0$        & $86 \pm 13 $ & $10.4 \pm 1.5 \pm 0.8$ & 
$0.03 \pm 0.36 \pm 0.09$ \\
\hline
$B^0\rightarrow\pi^0\pi^0$       & $23 \pm 10 $ & $<3.6 \;\; (1.6^{+0.7\; +0.6}_{-0.6\; -0.3})$  & \\
\hline
$B^+\rightarrow K^0\pi^+$        & $172 \pm 17$ & $17.5 \pm 1.8 \pm 1.3$ & 
$-0.17 \pm 0.10 \pm 0.02$ \\
\hline
$B^+\rightarrow K^0 K^+$         & $<10$        & $<1.3$                   & \\
\hline
\end{tabular}
\end{center}
\caption{Results of two-body branching fraction analyses\cite{babar_hphm,babar_hpi0,babar_hpi02,babar_2pi0}.  The $\pi^0\pi^0$ 
result has a statistical significance of $2.5\sigma$.  The results for
$K^0\pi^+$ and $K^0 K^+$ are based on 60 million $B{\bar B}$ pairs.
Upper limits are at the
90\% confidence level.}
\label{tab:2body}
\end{table}

Despite no central measurement of the $\pi^0\pi^0$ final state, 
it is still possible to
place limits on the relationship between the measured parameter $\alpha_{eff}$
and the Unitarity Triangle parameter $\alpha$.  Using the bound of Grossman and
Quinn\cite{grossman_quinn} and our measured values, we set an upper limit of
$|\alpha_{eff}-\alpha| < 51^{\circ} $ at 90\% CL.

\subsection{Quasi-Two-Body Decays}

Quasi-two-body decays proceed through resonant intermediate states.  The 
analysis of such modes is very similar to true two-body decays, but there are 
additional variables that provide separation between the signal and background,
such as the resonance invariant mass and polarization (if the final 
state is a pseudoscaler-vector combination).  We present the analyses of
three groups of related quasi-two-body decays; 
$B\rightarrow\phi K^{(*)}$\cite{babar_phi}, 
$B\rightarrow\omega h$\cite{babar_phi,babar_omega} and 
$B\rightarrow\eta^{(\prime)} K^{(*)}$\cite{babar_etap}.

\subsubsection{$B\rightarrow\phi K^{(*)}$}

\begin{figure}[t]
\begin{center}
%\leavevmode
\includegraphics[angle=0,keepaspectratio=true,height=1.5in]
{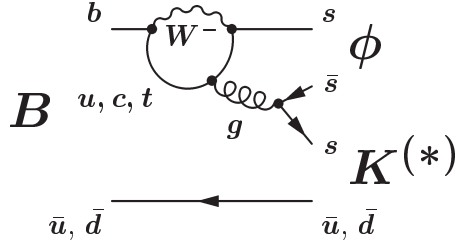}
\end{center}
%\vspace{-0.3in}
\caption{Leading diagram for the decay $B\rightarrow\phi K^{(*)}$.}
\label{fig:phi_diagram}
\end{figure}

The decay $b\rightarrow s{\bar s}s$ is CKM forbidden, thus the decay
$B\rightarrow\phi K^{(*)}$ is a nearly pure gluonic penguin, as shown in
Figure~\ref{fig:phi_diagram}.  New physics might not only manifest itself
as a deviation from the SM prediction for the decay rate, but since the mode
$\phi K^0_S$ is the strange analog to $J/\psi K^0_S$, the CP phase one measures
in a time-dependent analysis could be altered from it's SM value of $\beta$.

The signal yield in each of these modes is determined from an extended unbinned
maximum likelihood fit to $m_{ES}$, $\Delta E$, a Fisher discriminant and the 
$\phi\;(K^+K^-)$ invariant mass.  
For the $\phi K^0_S$, $\phi K^+$ and $\phi\pi^+$ 
final states, the $\phi$ polarization is included in the fit, as is the 
Cerenkov angle residual for the charged states.  For the $\phi K^*$ final
states, the $K^*$ invariant mass is included in the fit.
Significant signals are observed for both charged and
neutral $B$ decays to $K$ and $K^*$ final states.  The results of this
analysis on a sample of 60 million $B{\bar B}$ pairs are

\begin{eqnarray}
{\cal B}(B^+\rightarrow \phi K^+)
& = & (9.2 \pm 1.0 \pm 0.8) \times 10^{-6} \nonumber\\
 {\cal B}(B^+\rightarrow \phi K^0)
& = & (8.7^{+1.7}_{-1.5}  \pm 0.9) \times 10^{-6} \nonumber\\
{\cal B}(B^+\rightarrow \phi K^{*+})
& = & (9.7^{+4.2}_{-3.4} \pm 1.7) \times 10^{-6} \nonumber\\
{\cal B}(B^+\rightarrow \phi K^{*0})
& = & (8.7^{+2.5}_{-2.1} \pm 1.1) \times 10^{-6} \nonumber\\
{\cal B}(B^+\rightarrow \phi\pi^+) 
& < & 0.56 \times 10^{-6} \; @ \; 90\%\; {\rm CL}. 
\label{eqn20}
\end{eqnarray}
A stringent limit is also placed on the decay
$B^+\rightarrow\phi\pi^+$, which is both CKM and color suppressed.

\subsubsection{$B\rightarrow\omega h\;(h=K,\pi)$}

$B$ decays involving an omega and either a kaon or a pion proceed through
a mixture of CKM suppressed $b\rightarrow u$ tree and CKM forbidden 
$b\rightarrow d,s$ penguin amplitudes.  The analysis method is identical to 
that described for $\phi K^{(*)}$.  The results in Table~\ref{tab:omega} 
are based on a sample
of 22 million $B{\bar B}$, except for the $\omega K^0_S$ analysis, which 
was performed on a sample of 60 million $B{\bar B}$ pairs, and is a first 
observation.

\begin{table}[ht]
\begin{center}
%\hspace*{-1.5cm}
\begin{tabular}{|c|c|c|c|}
\hline
Final State & $N_{Signal}$ & $S(\sigma)$ & ${\cal B}\times 10^{-6}$ \\
\hline
\hline
$\omega K^+$ & $6.4^{+5.6}_{-4.4}$ & 1.3  & $<4\;(1.4^{+1.3}_{-1.0}\pm 0.3)$ \\
\hline
$\omega K^0$ & $26.6^{+7.7}_{-6.6}$ & 6.6 & $5.9^{+1.7}_{-1.5}\pm 0.9$ \\
\hline
$\omega \pi^+$&$27.6^{+8.8}_{-7.7}$ & 4.9 & $6.6^{+2.1}_{-1.8}\pm 0.7$ \\
\hline
$\omega \pi^0$&$-0.9^{+5.0}_{-3.2}$ & -   & $<3\;(-0.3\pm 1.1\pm 0.3)$ \\
\hline
\end{tabular}
\end{center}
\vspace{-0.5cm}
\caption{Results of the branching fraction analysis of 
$B\rightarrow \omega h$.  S is the statistic significance of the result.
Upper limits are at the 90\% confidence level.}
\label{tab:omega}
\end{table}

\subsubsection{$B\rightarrow\eta^{(\prime)} K^{(*)}$}

$B$ decays to $\eta$ and $\eta^\prime$ with a kaon or $K^*$ proceed 
predominantly through penguins, although there is some $b\rightarrow u$ tree
contribution as well.  The decays $B\rightarrow\eta^\prime K$ and 
$B\rightarrow\eta K^*$ were the first gluonic penguins to be 
observed\cite{cleo_etap}, and the rates are much larger than initially 
expected.  The best present conjecture\cite{etapk_theory} is that the tree
and penguin amplitudes interfere in such a way as to enhance $\eta^\prime K$
and $\eta K^*$ but suppress $\eta^\prime K^*$ and $\eta K$.  Because of it's
relatively large rate and nearly pure penguin content, 
$B\rightarrow\eta^\prime K^0_S$
is also of considerable interest for measurements of time-dependent CP 
asymmetries, which within the SM should probe the angle $\beta$.

Signals for these modes are extracted as described above for 
$\omega$ and $\phi$.  The $\eta^\prime$ is reconstructed in two decay chains;
$\eta(\gamma\gamma)\pi^+\pi^-$ and $\rho^0\gamma$.  The $\eta$ is reconstructed
as $\eta\rightarrow\gamma\gamma$ and $\eta\rightarrow\pi^+\pi^-\pi^0$.  The
results of these analyses are displayed in Table~\ref{tab:etap}.  The 
data samples used for the $\eta$ and $\eta^\prime$ analyses are 22 and 60
million $B{\bar B}$ pairs respectively.

\begin{table}[ht]
\begin{center}
%\hspace*{-1.5cm}
\begin{tabular}{|c|c|c|}
\hline
Final State & $N_{Signal}$ & ${\cal B}\times 10^{-6}$ \\
\hline
\hline
$\eta^\prime K^+$ & $445 \pm 26 $ & $67 \pm 5 \pm 5$ \\
\hline
$\eta^\prime K^0$ & $135 \pm 15 $ & $46 \pm 6 \pm 4$ \\
\hline
$\eta^\prime K^{*0}$ & $5.2 \pm 3.4 $ & $<13\;(4.0^{+3.5}_{-2.4}\pm 1.0)$ \\
\hline
$\eta K^+$ & $12.9 \pm 5.7 $ & $<6.4\;(3.8^{+1.8}_{-1.5}\pm 0.2)$ \\
\hline
$\eta \pi^+$ & $8.0 \pm 5.9 $ & $<5.2\;(2.2^{+1.8}_{-1.6}\pm 0.1)$ \\
\hline
$\eta K^0$ & $5.7 \pm 3.3 $ & $<12\;(6.0^{+3.8}_{-2.9}\pm 0.4)$ \\
\hline
$\eta K^{*0}$ & $20.5 \pm 6.3 $ & $19.8^{+6.5}_{-5.6}\pm 1.5$ \\
\hline
$\eta K^{*+}$ & $14.3 \pm 6.6 $ & $22.1^{+11.1}_{-9.2}\pm 3.2$ \\
\hline
\end{tabular}
\end{center}
\vspace{-0.5cm}
\caption{Results of the branching fraction analyses of 
$B\rightarrow \eta^{(\prime)} K^{(*)}$.  Upper limits are at the
90\% confidence level.}
\label{tab:etap}
\end{table}

\subsection{Three-Body Decays}

\begin{figure}[ht]
\begin{center}
%\leavevmode
\psfiletwo{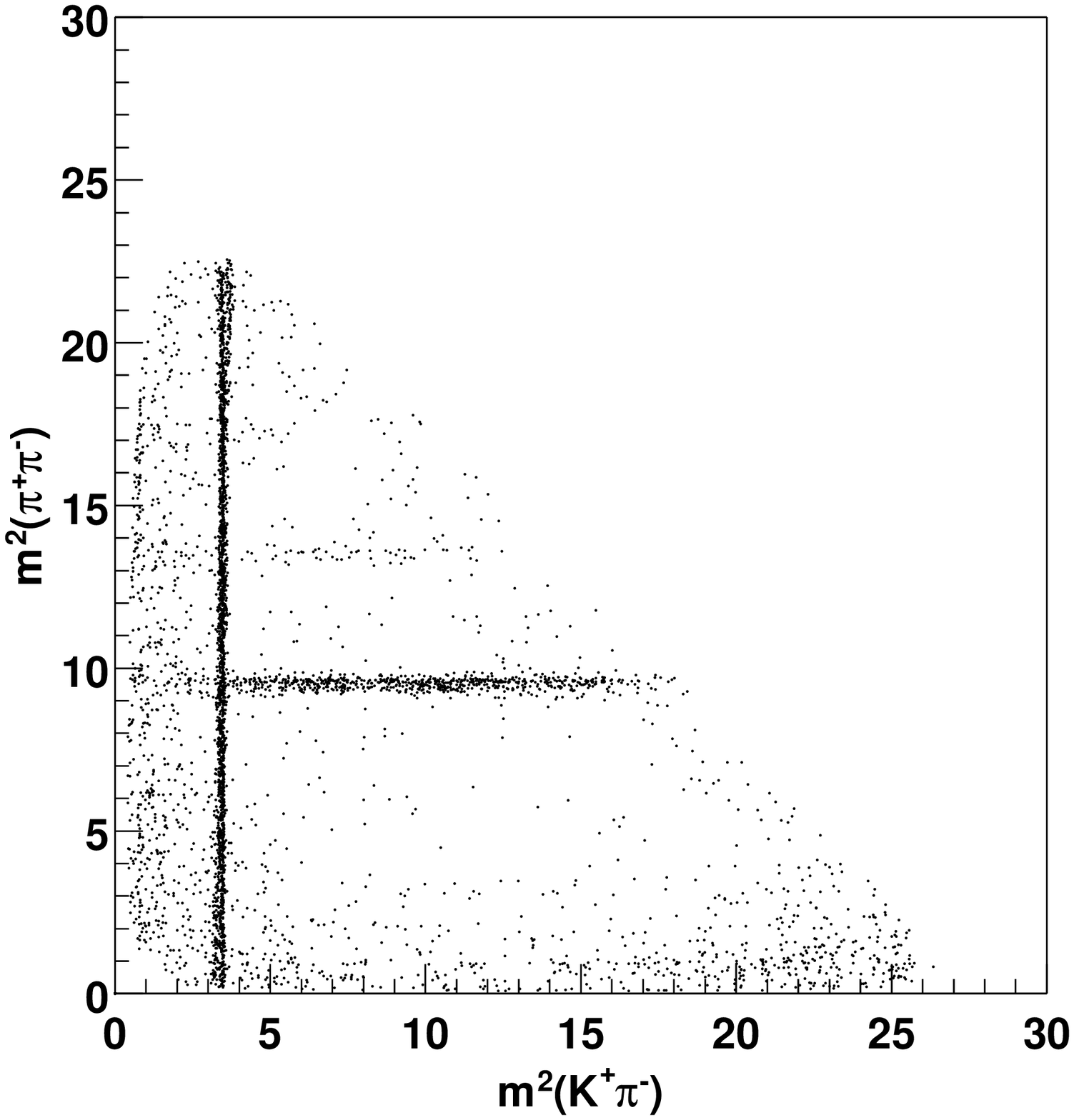}{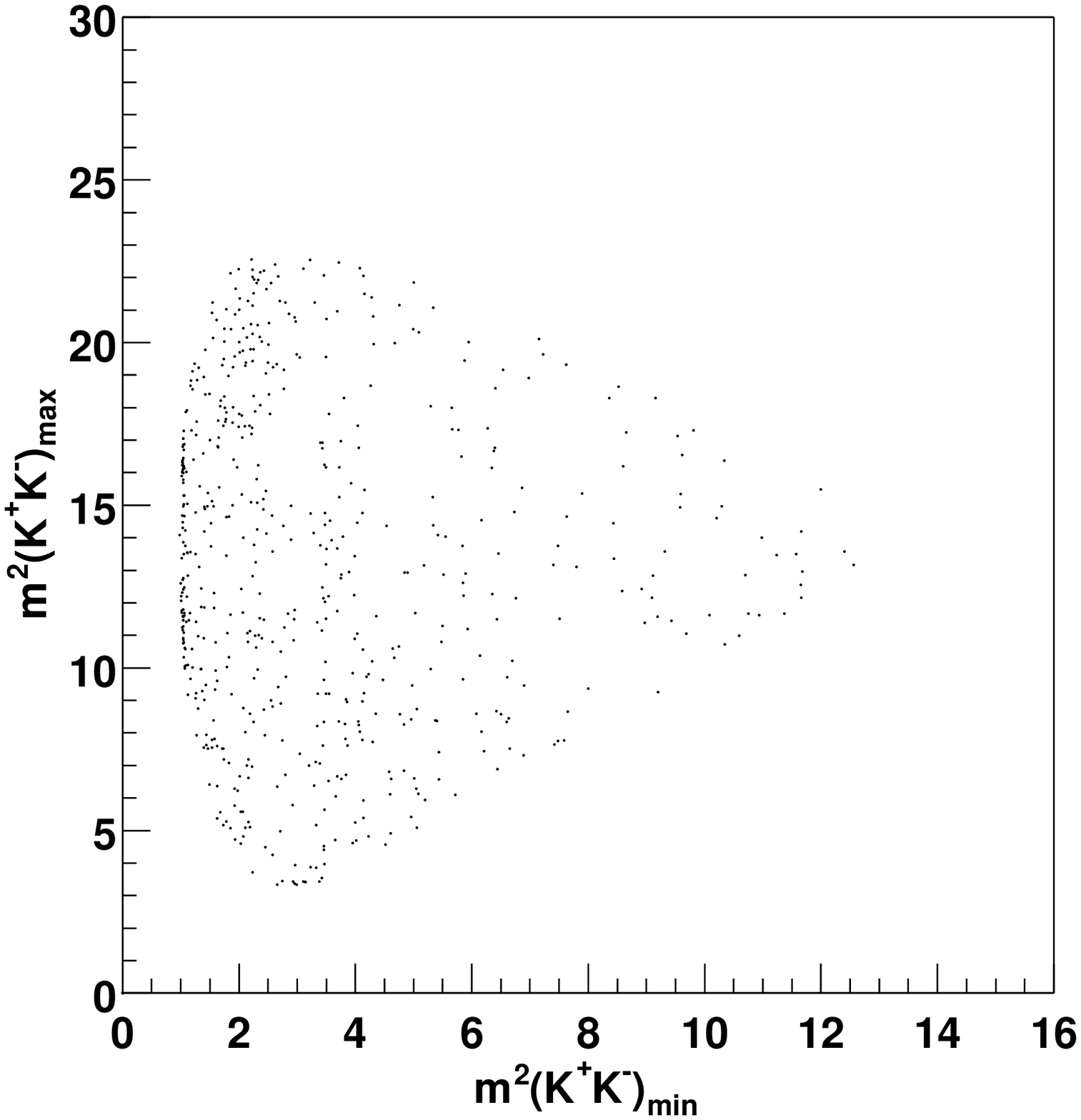}{1.0}%
\end{center}
\vspace{-1cm}
\caption{Unbinned dalitz plots for 
$K^+\pi^-\pi^+$ (left) and $K^+K^-K^+$ 
(right) for events in the 
signal region.  No efficiency corrections have been applied to the dalitz 
plots, and the charm contributions have not been removed.}
\label{fig:dalitz}
\end{figure}

We describe here the analysis of 
$B^+\rightarrow h^+h^-h^+$\cite{babar_3h}, 
where $h$ is either a charged kaon or pion.  An event counting analysis 
is performed over the full three particle dalitz plot.  All final states are
measured simultaneously, and unfolded to obtain branching fractions for each
combination.  Continuum background is suppressed using the thrust angle 
and a Fisher discriminant.  In addition to continuum background, the open 
nature of the dalitz plot also admits background in some regions from 
$B^+\rightarrow J/\Psi K^+$ and $B^+\rightarrow D\pi^+ / DK^+$. 
These regions of the dalitz plot are vetoed.  
Charged particle identification is 
crucial to this analysis, and along with tracking, is the primary source
of systematic uncertainty.  Figure~\ref{fig:dalitz} shows the dalitz plots
for $B^+\rightarrow K^+K^-K^+$ and 
$B^+\rightarrow K^+\pi^-\pi^+$.  Results of this analysis on a sample of 56 
million $B{\bar B}$ pairs are
\begin{eqnarray}
{\cal B}(B^{\pm}\rightarrow \pi^{\pm}\pi^{\mp}\pi^{\pm}) 
& < & 15 \times 10^{-6} \; @ \; 90\%\; {\rm CL} \; (8.5\pm 4.0\pm 3.6) \nonumber\\
{\cal B}(B^{\pm}\rightarrow K^{\pm}\pi^{\mp}\pi^{\pm}) 
& = & (59.2 \pm 4.7 (stat) \pm 4.9 (sys))\times 10^{-6} \nonumber\\
{\cal B}(B^{\pm}\rightarrow K^{\pm}K^{\mp}\pi^{\pm}) 
& < & 7 \times 10^{-6} \; @ \; 90\% \; {\rm CL} \; (2.1\pm 2.9\pm 2.0) \nonumber\\
{\cal B}(B^{\pm}\rightarrow K^{\pm}K^{\mp}K^{\pm}) 
& = & (34.7 \pm 2.0 (stat) \pm 1.8 (sys))\times 10^{-6}.
\label{eqn19}
\end{eqnarray}

\section{Conclusion and Outlook}

We have presented a number of results for rare $B$ meson 
decays using all or part of a sample of approximately 88 million $B{\bar B}$
pairs collected by the \babar\ detector.  Updates of many of these analyses
to the full data set are in progress.  These results represent only a part of 
the
spectrum of possible measurements of rare decays.  The larger data sets 
that will be available in the coming years will allow us to more fully 
exploit rare decays to test the self consistency of the flavor sector of the
Standard Model, and will perhaps offer the first glimpse of new physics which
lies beyond.

\section{Acknowledgments}

We are grateful for the excellent luminosity and machine conditions provided
by our PEP-II colleagues, and for the substantial dedicated effort from the 
computing organizations that support \babar. The author's work was performed
under the auspices of the U.S. Department of Energy by the University of
Colorado under Contract DE-FG03-95ER40894.


\begin{thebibliography}{10}
\bibitem{charge} Except as noted explicitly, we use a particle name to denote
either member of a charge conjugate pair.
\end{thebibliography}
\end{document}